\documentclass{emulateapj}
\usepackage{apjfonts}

\newcommand{\mjb}{mJy~beam$^{-1}$}
\newcommand{\jb}{Jy~beam$^{-1}$}
\newcommand{\kms}{km~s$^{-1}$}

\newcommand{\mum}{$\mu$m}
\newcommand{\cc}{cm$^{-3}$}

\newcommand{\tr}{$T_{rot}$}
\newcommand{\meth}{CH$_{3}$OH}
\newcommand{\so}{SO$_{2}$}
\newcommand{\hc}{HC$_{3}$N}
\newcommand{\cmw}{cm-wavelength}
\newcommand{\sfive}{$\sim -$5 km~s$^{-1}$}
\newcommand{\sten}{$\sim -$10.5 km~s$^{-1}$}

\shortauthors{Brogan et al.}
\shorttitle{Arcsecond Scale Complexity in Cepheus~A-East}

\begin{document}

\title{Arcsecond Scale Kinematic and Chemical Complexity in Cepheus~A-East}

\author{C.~L.\ Brogan\altaffilmark{1},
C.~J.\ Chandler\altaffilmark{1},
T.~ R.\ Hunter\altaffilmark{1}, 
Y.~L.\ Shirley\altaffilmark{2},
A.~P.\ Sarma\altaffilmark{3}}

\altaffiltext{1}{National Radio Astronomy Observatory; 
cbrogan@nrao.edu}

\altaffiltext{2}{University of Arizona; Bart J. Bok Fellow}

\altaffiltext{3}{Physics Department, DePaul University}

\begin{abstract}
    
We present results from SMA observations of the star forming region
Cepheus A-East at $\sim 340$~GHz (875~$\mu$m) with $0\farcs7 - 2''$
resolution. At least four compact submm continuum sources have been
detected, as well as a rich forest of hot core line emission. Two
kinematically, chemically, and thermally distinct regions of molecular
emission are present in the vicinity of the HW2 thermal jet, both
spatially distinct from the submm counterpart to HW2. We propose that
this emission is indicative of multiple protostars rather than a
massive disk as reported by \citet{Patel2005}.

\end{abstract}


\section{INTRODUCTION}
\label{sINTRO}

The Cepheus A-East (hereafter CepA) star forming region lies at a
distance $D\sim725$ pc and has $L_{\rm bol} \sim 2.2\times
10^4$ L$_{\sun}$, consistent with a cluster of B0.5 or later stars
(\citealt*{bhj59,sargent1979,mueller2002}).  At \cmw\/s, CepA consists
of several compact sources called HW1, HW2, ... HW9, which lie along a
roughly inverted Y-like structure \citep[e.g.,][]{hw84,garay96}.  It
is currently unclear how many of these compact ionized structures
correspond to individual protostars. For example, much of the \cmw\/
emission from the HW2 region is due to a bipolar thermal jet rather
than a Str\"omgren sphere (\citealt{rod94,Curiel2006}).

A wide range of other signposts of on-going star formation have been
observed in CepA including several outflow components
\citep[e.g.][and references therein]{codella2005}, however the
locations of the powering sources remain uncertain. On smaller
sizescales copious OH, H$_2$O, and CH$_3$OH maser emission has been
detected toward several of the cm-wavelength sources \citep[][and
references therein]{Vlemmings2006}. \citet{Patel2005} report the
detection of a massive molecular gas and dust disk toward the HW2
source from Submillimeter Array\footnote{The Submillimeter Array
is a joint project between the Smithsonian Astrophysical Observatory
and the Academia Sinica Institute of Astronomy and Astrophysics.} (SMA)
observations of methyl cyanide.

In this Letter, we present new and archival SMA 345~GHz observations
toward CepA \citep[including those of][]{Patel2005}, concentrating on
the continuum data and the spatial, kinematic, and temperature
information provided by the wealth of spectral lines; complete details
of the spectral line results will be presented in a future
paper. Our observations and results are presented in \S\
\ref{sOR}\, and are discussed in \S\ \ref{sDISC}.

\section{OBSERVATIONS \& RESULTS}
\label{sOR}

The SMA observing parameters are provided in Table~1.  The data were
taken with a channel width of 0.8125 MHz ($\sim 0.7$ \kms\/) except
for the archival data which have 8 times poorer spectral resolution
across most of the band (the CH$_3$CN (18--17) K=0-3 lines were
observed with $\sim 0.7$ \kms\/ resolution).  All five epochs employed
seven antennas. The data were calibrated using the MIRIAD software
package. The quasars BL~Lac and 3C~454.3 were used for phase
calibration. 3C~454.3, Uranus, and Saturn were used for bandpass
calibration. Comparison of the amplitude calibration (from the
quasars) applied to Uranus vs. a model of its baseline dependent flux
density suggests that the CepA amplitude calibration is accurate to
within $\sim 15\%$.

\begin{deluxetable}{lccll}
\tablecaption{SMA Observing Parameters}
\tablecolumns{5}
\tablehead{
\colhead{Date} & \colhead{$u$-$v$ range} 
& \colhead{t$_{\rm int}$} & \colhead{USB/LSB$^a$} & \colhead{Line Beam$^b$}\\ 
\colhead{} & \colhead{(k$\lambda$)} 
& \colhead{(hours)} & \colhead{(GHz)} &
\colhead{($\arcsec\times \arcsec$ ($\arcdeg$))} }
\startdata
30 Aug 2004$^c$ & 20-190  & 5.7 & 331.4/--    & $0.9\times0.8$ (77)\\ 
26 Sep, 18 Oct 2004  & 14-130  & 5.0 & 343.0/333.0 & $1.9\times1.2$ (71)\\ 
05, 07 Oct 2005     & 10-80   & 4.1 & 346.6/336.6 & $2.0\times1.9$ (33)  
\enddata
\tablenotetext{a} {Approximate center frequency of 2 GHz wide sidebands.}
\tablenotetext{b} {Synthesized beam and position angle of the USB line data.}
\tablenotetext{c} {Archival \citet{Patel2005} data, only the
  upper sideband was analyzed.}
\end{deluxetable}

Continuum subtraction, imaging, deconvolution, and
self-calibration were carried out in AIPS.  After extracting the
continuum using line-free channels in the $u$-$v$ data, the
continuum for each observation/sideband were separately
self-calibrated; the derived phase and amplitude corrections were also
applied to the line data sets. The spectral line data were
Hanning smoothed during imaging to produce a final spectral resolution
of $\sim 1.4$ \kms\/ (except for the archival SMA dataset with 5.9
\kms\/ spectral resolution). To create the highest possible
sensitivity continuum image, all of the final continuum datasets
were combined in the $u$-$v$ plane and imaged.

\subsection{CepA Submillimeter Continuum}

\begin{figure*}
\plotone{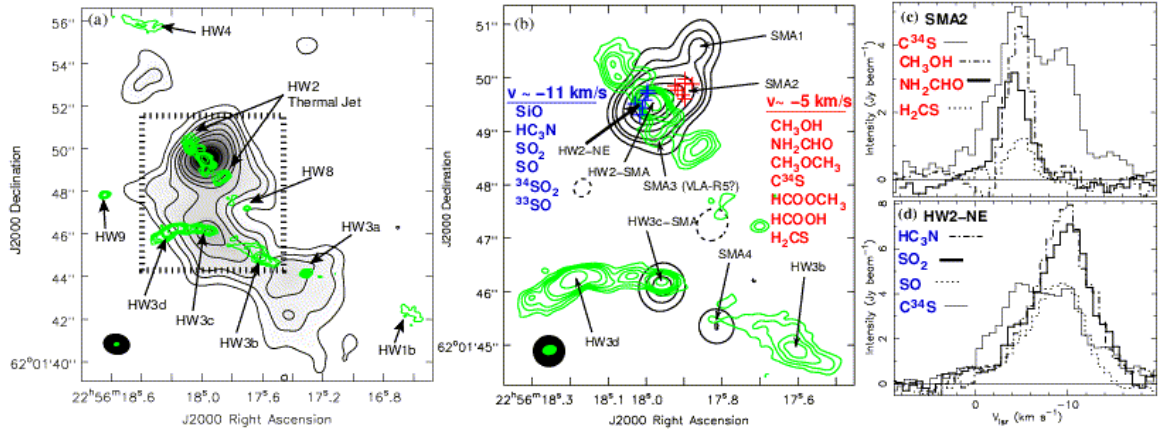}
\caption{(a) Combined SMA 875~$\mu$m continuum image (greyscale and
black contours) with a resolution of $1\farcs3\times 1\farcs0$
(P.A.=$79\arcdeg$) and contour levels of -40, 40 ($3\sigma$), 80, 120,
200, 300, 400, 600, 1000, and 1600 \mjb\/.  This image has not been
corrected for primary beam attenuation. Green VLA 3.6 cm contours at
0.06 ($3\sigma$), 0.12, 0.2, 0.3, 0.4, 0.5, and 1.5 \mjb\/ and
$0\farcs23$ resolution are superposed. The region shown in
Fig.~\ref{cont}b is indicated by the dashed box. (b) Superuniform
weighted 875~\mum\/ contour map (black) created using baselines longer
than 40 k$\lambda$ and restored with a $0\farcs6$ beam; the contour
levels are -50, 50 (4$\sigma$), 100, 150, 300, 500, 700, and 900
\mjb\/.  The green contours are the same as in Fig.~\ref{cont}a.  The
synthesized beams are shown in the lower left.  Colored crosses mark
the peak positions of the molecular species listed. Prominent
cm-$\lambda$ and submm sources are also labeled. Sample spectral line
profiles from the positions indicated on Fig. 1b for (c) SMA2 and (d)
HW2-NE. The displayed transitions are C$^{34}$S (7--6), CH$_3$OH
($14_{7,8}A^{\pm}$--$15_{6,9}A^{\pm}$), NH$_2$CHO
($16_{2,15}$--$15_{2,14}$), H$_2$CS ($10_{0,10}$--$9_{0,9}$), HC$_3$N
(38--37), SO$_2$ ($16_{7,9}$--$17_{6,12}$), and SO
($11_{10}$--$10_{10}$).  Note that at the spectral line angular
resolution of these transitions ($\sim 2\arcsec$) the two positions
are not completely independent.
\label{cont}}
\end{figure*}

Fig.~\ref{cont}a shows the combined naturally weighted 875~\mum\/ SMA
continuum image with a beam of $1\farcs3\times 1\farcs1$ and
integrated flux density of 7.7 Jy; the rms noise is 13 \mjb\/. The
peak of the single dish submm source \citep[see
e.g.][]{mueller2002} has been resolved for the first time into a
number of distinct components.  Archival 2002 Very Large
Array\footnote{The National Radio Astronomy Observatory operates the
Very Large Array and is a facility of the National Science Foundation
operated under a cooperative agreement by Associated Universities,
Inc.} (VLA) 3.6cm continuum contours are also shown on for comparison.
While there is rough agreement between the peak 875~\mum\/ continuum
emission and the well-known ionized thermal jet HW2
\citep[e.g.][]{rod94,garay96}, there is excess emission to the NW of
HW2 suggestive of additional unresolved structure.

In order to better compare the morphology of the cm and submm
emission, we have also created a superuniform weighted image using
baselines $>40~{\rm k}\lambda$, resulting in a beam size of
$0\farcs8\times0\farcs7$. To further delineate the morphology of the
emission in the vicinity of HW2, in Fig.~\ref{cont}b we present the
superuniform weighted image restored with a $0\farcs6$ beam
(equivalent to 1.8 times the longest baseline sampled) which
emphasizes the locations of the clean components.  Fig.~\ref{cont}b
reveals the presence of at least two distinct submm sources in the
vicinity of HW2, HW2-SMA and SMA1, in addition to an extension NW of
HW2-SMA which we denote SMA2 (see \S3).  A weak extension south of
HW2-SMA (denoted SMA3) is also visible, which may be a submm
counterpart to the low mass protostar VLA-R5 \citep{Curiel2002}. The
combined morphology of HW2-SMA, SMA2, and SMA3 is the structure
reported by \citet{Patel2005} as a dust disk; their Fig.~1 with $\sim
0\farcs75$ resolution also shows that the centroid of submm emission
is not centered on the axis of the HW2 jet.  The HW2-SMA 875 \mum\/
continuum peak is within $0\farcs1$ of the proposed location of the
powering source of the HW2 thermal jet based on high resolution
($0\farcs05$) 7 mm VLA data \citep{Curiel2006} which is well within
our absolution position uncertainty of $0\farcs15$. Two additional
submm cores are detected to the south of HW2, one coincident with the
cm-wavelength source HW3c (denoted HW3c-SMA) and another located at
the NE tip of the cm-wavelength source HW3b (designated SMA4). No
distinct compact 875~\mum\/ counterparts to HW8, HW9, HW3a, HW3b or
HW3d are detected in the SMA data.

\subsection{CepA Hot Core Line Emission}

Within the 8 GHz of total bandwidth observed (Table~1) we detect more
than 20 distinct species along with a number of their isotopologues.
A few transitions with low excitation energies ($^{12}$CO(3--2),
CS(7--6), H$^{13}$CO$^+$(4--3)) show extended emission over the full
primary beam ($\sim 36\arcsec$) corresponding to the previously
studied outflows \citep[see e.g.][]{codella2005}.  However, the
emission from most species is restricted to compact regions ($\lesssim
2\arcsec$) that coincide with the submm continuum emission
(Fig.~\ref{cont}b).  At $1\arcsec-2\arcsec$ resolution (Table 1) the
species exhibiting compact emission in the vicinity of HW2 are
strongest at one of two distinct velocities: $-5.0\pm0.5$ or
$-10.5\pm0.5$ \kms\ \citep[Fig.~\ref{cont}b,c,d; see
also][]{codella2006}. These two kinematic features are also spatially
distinct. Without exception, molecules that are strongest at \sten\/
have peak positions $\sim0\farcs25$ (200 AU) E/NE of HW2-SMA (we denote
this position HW2-NE), and molecules that are strongest at \sfive\/
peak at the position of SMA2.  Although this dichotomy exists for all
observed species, a few abundant high density tracers like CH$_3$CN
and C$^{34}$S, show emission of nearly equal strength toward both
positions (Fig.~\ref{cont}c,d).

Using 3mm Plateau de Bure (PdBI) data \citet{MartinPintado2005} also
find that \so\/ emission peaks to the E/NE of HW2 at a velocity of
\sten\/ and suggest that this emission is due to a distinct
intermediate mass protostar; the SMA and PdBI \so\/ positions agree to
within the absolute position uncertainty of $0\farcs15$.  This result
has recently been confirmed by the VLA detection of \so\/ emission and
weak 7mm continuum emission at the position of HW2-NE
\citep{Jimenez2007}.  In an extensive PdBI spectral line study Comito
et al. (in preparation) also find strong spatial, chemical, and
kinematic differentiation in general agreement with the SMA results.
The two velocity components at \sfive\/ and \sten\/ have also been
observed in single dish H$_2$CS, CH$_3$OH, and HDO data with
resolutions ranging from $10''$ to $30''$ \citep{codella2006}.  The
single dish emission from both velocity components is extended.  Our
SMA data are insensitive to spatial structures $\gtrsim 15\arcsec$,
but it seems likely that SMA2 is associated with the larger scale
\sfive\ component.

\begin{figure}
\epsscale{1.0}
\plotone{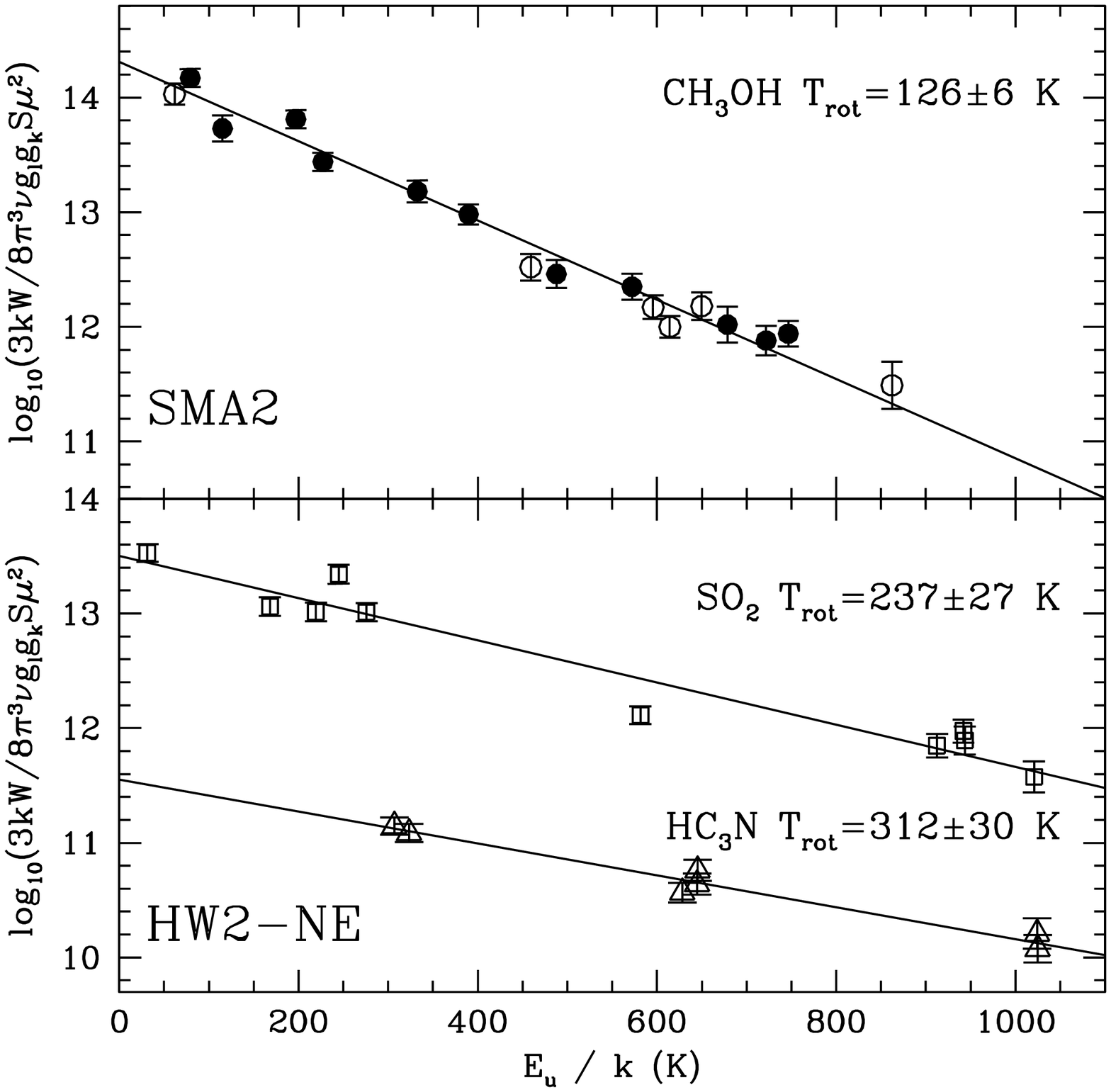}
\caption{Rotation diagrams for \meth\/ A ($\bullet$) and E ($\circ$)
transitions toward SMA2 and for \so\/ ($\sq$) and 
\hc\/ ($\bigtriangleup$) toward HW2-NE (see Fig. 1b). The
\meth\/ and \so\/ data have been corrected for optical depth
effects. 
\label{rot}}
\end{figure}

For \meth\/ towards SMA2 (at \sfive\/), and \so\/ and \hc\/ toward
HW2-NE (at \sten\/), we have measured enough transitions to construct
rotation diagrams \citep[Fig.~\ref{rot}; see][]{Goldsmith1999}.
Emission from the two velocity components are well-separated, even
though they are not completely spatially resolved by our observations
(Fig.~1c,d).  The \meth\/ and \so\/ data have been corrected for
optical depth effects by first estimating the opacity of one
transition using a less abundant isotopologue and assuming Galactic
abundance ratios of $^{12}$C/$^{13}$C=70 and $^{32}$S/$^{34}$S=23
\citep[][and references therein]{Milam2005,Chin1996}. Then, assuming
LTE, the other transitions were corrected using the formalism
described in \citet{Sutton2004}.  The \meth\/
optical depths are significant (up to 28.2); the \so\/ optical depth
is more moderate (up to 2.4). A moderate optical depth in the lower
lying \hc\/ lines is also possible \citep[see e.g.][]{wyrowski1999},
correction for which would yield an \hc\/ \tr\/ more consistent with
\so\/. Three of the lower energy A-type transitions of \meth\/ and the
two lowest energy transitions of \hc\/ are also detected toward
HW3c-SMA and SMA4; rotation diagram analysis for these two submm cores
yield \tr\/=$65\pm 25$ K.

In addition to the kinematic and chemical dichotomy between SMA2
and HW2-NE, their \tr\ differ by more than 100 K\@.
Both values of \tr\ (Fig. 2) are significantly larger
than reported by \citet{Torrelles1999} based on $\sim 1\arcsec$
resolution VLA NH$_3$ data (\tr\/=30-50 K), influenced by their
reported non-detection of NH$_3$ (4,4) [$E_{u}/k=202$ K].  An NH$_3$
(4,4) integrated intensity image from our reduction of these archival
data is shown in Figure 3; emission is clearly detected toward both
HW2-NE and SMA2.  A new analysis of the NH$_3$ data gives \tr\
within a factor of two of those reported here.  Our \tr\/ are also
larger than those estimated by \citet{Patel2005} (25-75 K) though the
sense of the temperature gradient is in agreement (i.e. warmer in the
E than W). Although \citet{Patel2005} report that CH$_3$CN (18-17) is
not detected above K=3, our reduction of their archival SMA data
detects CH$_3$CN up through K=8 ($E_{u}/k=607$ K, K=6 is shown in
Fig.~3), and we find similar \tr\/ to those shown in Fig.~2. Using 30m
data, \citet{MartinPintado2005} derived \tr\/$\sim 150$ K for 3mm
\so\/ and HC$_3$N transitions toward HW2-NE, however, beam dilution
may well play a role in this single dish result.

\section{DISCUSSION}
\label{sDISC}

\begin{figure*}
\epsscale{1.0}
\plotone{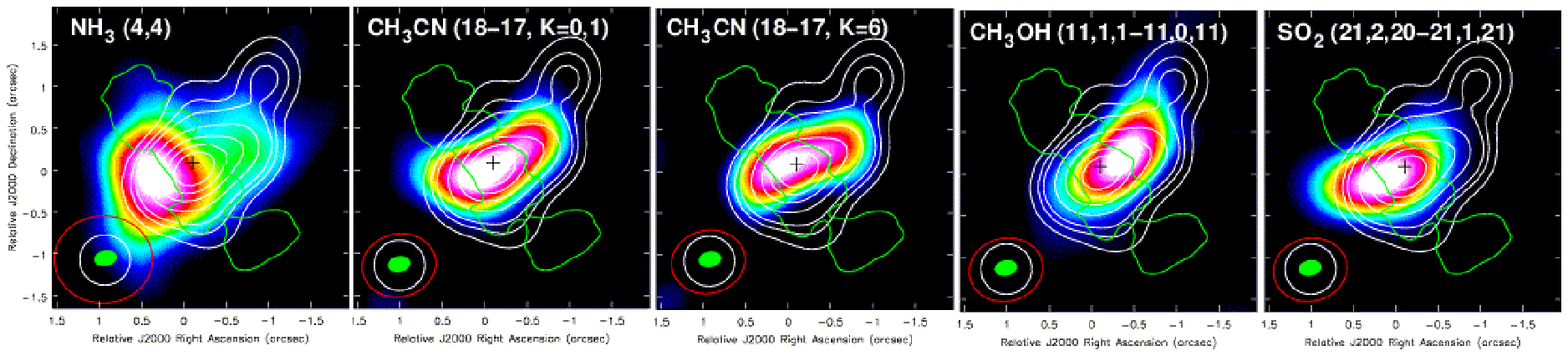}
\caption[f3.eps]{Integrated intensity images (over the full width of
the observed emission) of archival VLA NH$_3$ (4,4) data with
$1\farcs1\times 1\farcs0$ resolution and archival SMA data with
$0\farcs87\times0\farcs75$ resolution.  White 875 \mum\/ contours from
Fig. 1b, and a green 0.6 \mjb\/ 3.6cm contour are superposed.  The
black $+$ symbol shows the proposed location of the star powering the
HW2 thermal jet \citep{Curiel2006}. The beams are shown in the lower
left and the reference position is $22^{\rm h}56^{\rm m}17.998^{\rm
s}$, 62$\arcdeg$01$\arcmin$49$\farcs50$ (J2000).
\label{compare}}                                     
\end{figure*}

\begin{deluxetable}{llcccc}
\tablecaption{Derived Parameters of Submillimeter Cores}
\tablecolumns{6}
\tablehead{
\colhead{Name} & \colhead{R.A., Dec.$^a$} & \colhead{$S_{\nu}^b$} 
& \colhead{$T_b$} & \colhead{$T_d$} 
& \colhead{$M_{gas}$} \\ 
\colhead{} & \colhead{(s, $\arcsec$)} & \colhead{(\jb\/)} 
& \colhead{(K)} & \colhead{(K)} 
& \colhead{($M_\sun$)} }
\startdata
 SMA1     & 17.873, 50.56 & 0.28  & 5.0  &  20-100  & 1.0-0.1 \\ 
 SMA2     & 17.927, 49.74 & 0.80  & 14.4 & 115-135 & 0.32-0.27 \\
 HW2-SMA  & 17.999, 49.43 & 1.17  & 21.0 &  40-100 & 2.0-0.6 \\ 
 HW3c-SMA & 17.949, 46.07 & 0.25  & 4.5  &  40-90  & 0.3-0.1 \\  
 SMA4     & 17.820, 45.33 & 0.16  & 2.9  &  40-90  & 0.2-0.1 
\enddata
\tablenotetext{a} {J2000 coordinates added to 22h56m, 62$\arcdeg$01$\arcmin$,
the relative and absolute position uncertainties are better than $0\farcs02$ 
and $0\farcs15$, respectively.}
\tablenotetext{b} {Using superuniform weighted image with restoring 
beam $0\farcs84\times0\farcs70$.}
\end{deluxetable}

The observed distribution of submm continuum emission in the immediate
vicinity of HW2 ($\pm 0.5''$) allows for two possible interpretations:
a single elongated structure or two (or more) marginally-resolved
individual sources.  Although the kinematic dichotomy between the
positions labeled HW2-NE and SMA2 (Fig.~1b) could be interpreted as a
velocity gradient across a continuous structure
\citep[e.g.][]{Patel2005}, the dramatic chemical and thermal
differentiation demonstrated by our multi-species analysis is
difficult to explain with this picture.  Chemical and thermal
gradients might be expected in the radial and vertical directions
within a protostellar disk, but such dramatic azimuthal asymmetries
are more difficult to conceive in a single source scenario.  We also
note that beyond a velocity gradient, the kinematic evidence for a
Keplerian rotating disk is quite weak. For example, the weakness of
the emission at the central HW2 position in position-velocity (P-V)
diagrams such as those in \citet[][and in our own data, but not
shown]{Patel2005} are inconsistent with theoretical expectations
unless a central hole is invoked \citep[e.g.][]{Richer1991}.  Even
with this modification, it remains difficult to explain the unequal
position offsets of the two velocity peaks from the HW2 stellar
position \citep[see NH$_3$ in Fig.3 and the K=3 P-V diagram
of][]{Patel2005}.  In contrast, the presence of multiple sources at
different velocities naturally explains the observed behavior.  We
therefore favor the multiple source hypothesis \citep[see
also][]{MartinPintado2005}, with at least three sources in the
vicinity of HW2 (HW2-SMA, HW2-NE, and SMA2).  The remaining discussion
proceeds with this interpretation.

Table 2 summarizes the properties of the submm cores identified in
Fig.~1b (excluding SMA3).  The gas masses were estimated using
\begin{equation}
M_{gas}=\frac{R~S_{\nu}~D^2~\tau_{dust}}{B[\nu,T_d]~\kappa(\nu)~(1-{\rm
    exp}[-\tau_{dust}])},
\end{equation}
and assuming a gas-to-dust ratio $R=100$, $D=725$
pc, and a dust opacity $\kappa_{875 \mu m}=1.84$ cm$^2$
g$^{-1}$ extrapolated from \citet{Ossenkopf1994} for thin ice
mantles and density $10^6$ \cc\/. Values for the peak flux density
$S_\nu$ (none of the cores appear to be resolved), the continuum
brightness temperature ($T_b$), and the range of assumed dust
temperatures ($T_d$) are also listed in Table~2.  The $M_{gas}$ have
been corrected for the continuum opacity calculated from
$\tau_{dust}=-{\rm ln}[1-(T_b/T_d)]$.

The derived masses are very sensitive to the assumed temperatures
(Table 2). For SMA2, HW3c, and SMA4, we have used the range of dust
temperatures from rotation diagram analysis (\S 2.2). Temperatures for
HW2-SMA and SMA1 are difficult to estimate due to the proximity of
strong emission from HW2-NE and SMA2. No species peak at the position
of HW2-SMA suggesting it is not very warm (i.e. compared to SMA2 or
HW2-NE), though its association with the strong bipolar jet suggests
it is unlikely to be very cold either. Thus we have assumed a moderate
temperature range of 40 to 100 K for HW2-SMA. Since SMA1 is lacking
both a cm-wavelength counterpart and any distinct line emission we
assume a cooler lower limit (20 K).  With these assumptions the total
gas masses are fairly low, ranging from 0.1 to 2.0 M$_{\sun}$.  If
these are preprotostellar cores, this result implies that a {\em
massive} star will not form; if, however, an embedded star is present
(almost certainly true for HW2-SMA) these masses are consistent with
typical values observed for intermediate mass protostellar disks
\citep[e.g.][]{Hamidouche2006}.
 
No compact submm emission corresponding to HW8, HW9, HW3a, HW3b, or
HW3d is detected. HW3a, HW8 and HW9 are variable at \cmw\/s and are
thought to be low mass pre-main-sequence stars (e.g.,
\citealt{Hughes1988,garay96}), hence it is unsurprising that any submm
emission is below our $3\sigma$ detection threshold of 40 \mjb\/.
Based in part on the detection of very compact ($0\farcs1$) 2 cm
emission toward HW3d \citep{Hughes1988}, \citet{garay96} suggest that
HW3d contains its own internal energy source but also has thermal
jet-like extended \cmw\/ emission.  Given its lack of distinct submm
continuum or line emission we suggest instead that HW3d is a one-sided
ionized jet emanating from HW3c-SMA (see Fig. 1b). HW3b is consistent
with being a one-sided jet emanating from SMA4 or possibly the
counterjet to HW3d \citep[also see][]{garay96}.

We detect at least five submm sources (HW2-SMA, SMA1, SMA2, HW3c-SMA,
and SMA4) within a projected radius of $4\arcsec$ (2900~AU). If the
five low mass sources HW3a, HW8, HW9, VLA-R5, and VLA-R4
\citep{Hughes1988,Curiel2002} are included and equal clustering in the
perpendicular dimension is assumed, then the implied protostellar
density is $5.7 \times 10^5~{\rm pc}^{-3}$.  This approaches the
minimum theoretical value ($10^6~{\rm pc}^{-3}$) needed to test the
induced binary merger hypothesis proposed as a formation mechanism for
the most massive stars \citep{Bonnell05}.

\acknowledgments

We thank the SMA staff for their assistance.  This research used the
JPL (http://spec.jpl.nasa.gov) and Cologne
(http://www.ph1.uni-koeln.de/vorhersagen) molecular spectroscopy
databases.  This work has been partially supported by start-up funds
to A.P.S.\ at DePaul Univ.

\end{document}